

\null
\magnification 1200


\newskip\ttglue

\font\eightrm=cmr8
\font\eighti=cmmi8
\font\eightsy=cmsy8
\font\eightbf=cmbx8
\font\eighttt=cmtt8
\font\eightsl=cmsl8
\font\eightit=cmti8
\font\sixrm=cmr6
\font\sixbf=cmbx6
\font\sixi=cmmi6
\font\sixsy=cmsy6

\def \eightpoint{\def\rm{\fam0\eightrm}
\textfont0=\eightrm \scriptfont0=\sixrm \scriptscriptfont0=\fiverm
\textfont1=\eighti \scriptfont1=\sixi   \scriptscriptfont1=\fivei
\textfont2=\eightsy \scriptfont2=\sixsy   \scriptscriptfont2=\fivesy
\textfont3=\tenex \scriptfont3=\tenex   \scriptscriptfont3=\tenex
\textfont\itfam=\eightit  \def\it{\fam\itfam\eightit}%
\textfont\slfam=\eightsl  \def\sl{\fam\slfam\eightsl}%
\textfont\ttfam=\eighttt  \def\tt{\fam\ttfam\eighttt}%
\textfont\bffam=\eightbf  \scriptfont\bffam=\sixbf
 \scriptscriptfont\bffam=\fivebf  \def\bf{\fam\bffam\eightbf}%
\tt \ttglue=.5em plus.25em minus.15em
\setbox\strutbox=\hbox{\vrule height7pt depth2pt width0pt}%
\normalbaselineskip=9pt
\let\sc=\sixrm  \let\big=\eightbig  \normalbaselines\rm
}
\def\a{\alpha}

\def\d{\delta}

\def\h{\eta}

\def\l{\lambda}
\def\m{\mu}
\def\n{\nu}

\def\r{\rho}
\def\s{\sigma}
\def\t{\tau}

\def\C{\Gamma}



{\nopagenumbers

 \line{\hfil SWAT 95/70   }
\vskip2cm
\centerline{\bf ``FASTER THAN LIGHT'' PHOTONS IN GRAVITATIONAL FIELDS}
\vskip0.8cm
\centerline{\bf --~CAUSALITY, ANOMALIES AND HORIZONS}
\vskip1.5cm
\centerline{\bf G.M. Shore}
\vskip0.8cm
\centerline{\it Department of Physics}
\centerline{\it University of Wales Swansea}
\centerline{\it Singleton Park}
\centerline{\it Swansea, SA2 8PP, U.K. }
\vskip1.5cm
\noindent{\bf Abstract}
\vskip0.5cm
\noindent A number of general issues relating to superluminal photon
propagation in gravitational fields are explored. The possibility
of superluminal, yet causal, photon propagation arises because of
Equivalence Principle violating interactions induced by vacuum
polarisation in QED in curved spacetime. Two general theorems
are presented: first, a polarisation sum rule which relates the
polarisation averaged velocity shift to the matter energy-momentum
tensor and second, a `horizon theorem' which ensures that the
geometric event horizon for black hole spacetimes remains a true
horizon for real photon propagation in QED. A comparision is made
with the equivalent results for electromagnetic birefringence
and possible connections between superluminal photon propagation,
causality and the conformal anomaly are exposed.

\vskip2cm
\line{SWAT 95/70     \hfil}
\line{April 1995     \hfil}

\vfill\eject }

\vsize 19.5cm
\pageno = 1

\noindent {\bf 1. Introduction}
\vskip0.5cm

The possibility of superluminal photon propagation in gravitational
fields is one of the most remarkable predictions of quantum field
theory in curved spacetime. It appears that real photons propagating
in a variety of background spacetimes may, depending on their
direction and polarisation, travel with speeds exceeding the
normal speed of light $c$.

This phenomenon was discovered by Drummond and Hathrell in 1980 [1].
It is a quantum effect induced by vacuum polarisation and implies
that the Principle of Equivalence is violated in interacting
quantum field theories such as QED.

In their original paper [1], Drummond and Hathrell studied photon
propagation in Schwarzschild, Robertson-Walker, gravitational wave
and de Sitter backgrounds. In each case, except the totally isotropic
de Sitter spacetime, it was possible to find directions and
polarisations for which the photon velocity exceeds $c$.
In a subsequent paper with Daniels[2], we extended this analysis to the
Reissner-Nordstr\"om spacetime describing a charged black hole
with similar results. Previously, Ohkuwa[3] had generalised the
Drummond-Hathrell result to massless neutrino propagation in
a Robertson-Walker metric using the Weinberg-Salam model.

The effect is best described as a modification of the light cone
in a local inertial frame (LIF) to
$\bigl(\h_{ab} + \a \s_{ab}(R)\bigr) k^a k^b = 0$,
where $\h_{ab}$ is the Minkowski metric, $\a$ is the fine structure
constant and $\s_{ab}(R)$ depends on the Riemann curvature at the
origin of the LIF.
The correction arises from vacuum polarisation induced interactions
in the QED effective action of the typical form
$\a {1\over m^2} R_{\m\n\l\r} F^{\m\n} F^{\l\r}$,
where $m$ is the electron mass.
Such direct curvature couplings violate the Principle of
Equivalence. As discussed in section 2, it is this fact which allows
spacelike motion in a LIF without necessarily implying a violation
of causality.

The physical origin of this effect may be understood in qualitative
terms as follows. Vacuum polarisation allows the photon to exist
as a virtual $e^+ e^-$ pair and so at the quantum level
it is characterised by a spacetime scale $\l_c$, the Compton
wavelength of the electron. In a gravitational field, the photon
is therefore sensitive to an anisotropic spacetime curvature
and its characteristics of propagation may become curvature
dependent. The spatial anisotropy of the gravitational field results
in a polarisation dependence of the effect
(gravitational birefringence).

\vskip0.3cm
In this paper, we collect a number of new results and theorems
on the basic interpretation of the superluminal effect and its
realisation in black hole spacetimes.

\vskip0.3cm
The Principle of Equivalence would require that the only allowed
kinetic term in the effective action for the electromagnetic
field is simply $F_{\m\n} F^{\m\n}$. In section 2, we review the
geometric optics approximation for photon propagation and show
how this requirement implies that photon trajectories are null
geodesics. We then include the new vacuum polarisation induced
interactions and derive the modified, curvature-dependent equations
for the photon trajectories.

The violation of the Principle of Equivalence allows the possibility
of superluminal propagation without causality violation and we discuss
the conditions that must be satisfied to maintain causality.
We also discuss the intrinsically quantum question of the propagation
of photons whose polarisations do not satisfy the curvature
dependent equation of motion.

One of the main potential applications of superluminal propagation is
to the physics of black holes. The case of the
Schwarzschild black hole was discussed in ref.[1] and the extension
to charged (Reissner-Nordstr\"om) black holes was given in our
previous paper[2]. The analogous results for the Kerr metric describing
a rotating black hole will be presented in ref.[4]. We again find
a variety of directions and polarisations for which the photons have
superluminal velocities. Unlike the Schwarzschild and
Reissner-Nordstr\"om spacetimes, however, radially directed photons
in the Kerr metric have velocities differing from $c$, except at the
event horizon. It turns out that this is a non-trivial special case
of a general `horizon theorem', which is stated precisely
and proved here in section 3. The theorem states that in a general
spacetime with an event horizon, the light cone for photons travelling
normal to the horizon surface remains $k^2 = 0$ even in the
presence of the vacuum polarisation induced interactions.
The physical implications of this theorem merit further investigation,
but the result seems to ensure that the geometric event horizons
of the classical spacetime continue to be true horizons for real
photon trajectories.

An amusing application of this theorem is to de Sitter spacetime.
This has a cosmological event horizon, specific
to each observer. In fact, every point in the spacetime lies
on the horizon of some observer and thus, assuming the validity
of the horizon theorem in the cosmological context,
the velocity of light must be $c$ everywhere. This is in agreement with
the conclusion that the light cone is unchanged because the
spacetime is totally isotropic.

\vskip0.3cm
A new physical insight into the more general phenomenon of photon
propagation in non-trivial vacua has been given by Latorre, Pascual and
Tarrach[5]. Modifications to the velocity of light occur in a variety of
situations apart from gravitational backgrounds, including propagation
in electromagnetic background fields[6,2], Casimir-type regions with
boundaries[7,8,9], finite temperature[5] or density, etc. Latorre et
al.~have identified an intriguing general formula covering all these
cases, relating the polarisation (and, if necessary, direction)
averaged velocity shift to the background energy density, with a
universal numerical coefficient.\footnote{$^*$}{\eightpoint
\noindent Curiously, the coefficient involves the unusual number 11,
hinting at a possible relation with the conformal anomaly.
The numerical coefficient of the Euler term in the
gravitational conformal anomaly is $N_S + 11 N_F + 62 N_V$, where
$N_S, N_F$ and $N_V$ are the number of scalar, fermion and vector
fields respectively. In the conformal anomaly, the factor 11
arises from a one-loop, background field calculation involving
a single fermion, as does the effective action required to calculate
corrections to photon propagation.}
In cases where the background energy is positive, the average velocity
is less than $c$ whereas in negative energy situations such as
Casimir regions, the average velocity is greater than $c$.

Electromagnetic birefringence, described by Adler in ref.[6], was
studied for an arbitrary anisotropic (but homogeneous) background
electromagnetic field in our earlier paper[2]. In section 4, we make
some further observations. First, motivated by Latorre et al.,
we rewrite our previous result for the photon velocities in such a way
as to make clear that, for an arbitrary background, the direction and
polarisation averaged photon velocity shift is indeed proportional
to the electromagnetic field energy, with the required numerical
coefficient.

Superluminal propagation in this Minkowski spacetime situation
would violate causality and so the velocity shifts for both
polarisations must be negative. This requires certain combinations
of coefficients appearing in the effective action to be negative.
Remarkably enough, we find that this is precisely the condition
for the VEV of the trace of the energy-momentum tensor in the
electromagnetic background to be negative, although the physical
significance of such a requirement is not immediately clear.
This is another piece of evidence suggesting a deeper connection
between the conformal anomaly and photon propagation. In a sense,
some such relation might be expected since the presence in the quantum
theory of background interaction terms in the effective action
necessarily introduces a mass scale, removing the conformal invariance
of the classical photon action.

The explicit results for Ricci flat metrics, including Schwarzschild
and Kerr, suggest a second theorem, which we prove in section 3.
We show that in Ricci flat spacetimes, the velocity shifts
are equal and opposite in sign for the two physical
transverse polarisations. The polarisation averaged velocity shift
is therefore zero in these vacuum solutions of the Einstein equation
where the matter energy-momentum tensor vanishes.
However, the fact that a superluminal birefringent effect exists
even for Ricci flat spacetimes (with $T_{\m\n} = 0$) shows that
vacuum energy density cannot provide the whole explanation for
modified photon propagation. Birefringence associated with an
explicit Riemann curvature interaction in a non-isotropic
gravitational field is an alternative mechanism to modify the
light cone.

The generalisation of this polarisation sum rule for non Ricci flat
spacetimes is also given in section 3. This is particularly relevant
for the Friedmann-Robertson-Walker and Reissner-Nordstr\"om spacetimes.
In the general case, the polarisation average acquires a term
proportional to
$T_{\m\n}e^\m e^\n$, where $e^\m = k^\m/|\underline k|$ and $T_{\m\n}$
is the matter energy-momentum tensor, in accord with the specific
results of refs.[1] and [2]. This term is precisely the
same as arises in the case of electromagnetic birefringence.
The additional Riemann curvature term which induces birefringence
in the gravitational case is identified as the Newman-Penrose
scalar $\Psi_0$.

\vskip0.3cm
The major interpretational issues left open in previous work have
concerned the question of dispersion and whether this superluminal
propagation is observable in principle.
These difficulties primarily concern the specific mechanism of
generating explicit curvature couplings through vacuum polarisation
proposed in ref.[1].
We address the problem of dispersion in the context of the effective
action for QED in a background gravitational field in ref.[10]. It
should be emphasised, however, that the interest in such Equivalence
Principle violating interactions is broader than this. Once we have
established that such interactions may exist without necessarily
involving causality violation, it becomes an experimental question
whether they do in fact exist and with what characteristic length scale.
Such interactions could be searched for, for example, by studying
polarisation dependence in gravitational lensing.\footnote{$^*$}
{\eightpoint \noindent This suggestion is due to I.T. Drummond
(private communication). }

Of course, for weak fields the magnitude of the vacuum polarisation
induced effect is tiny. For example, the modification to the angle
of deflection of light by the sun is only a factor of $O(10^{-47})$ [1].
However, this is typical of all quantum field effects in macroscopic
gravitational fields, an important example being Hawking radiation[11].
Just as for Hawking radiation, the superluminal effect should
become large for curvatures comparable with the quantum scale, in this
case $\l_c$. Such curvatures may arise either for microscopic
black holes or in the very early universe.

\vfill\eject

\noindent{\bf 2. Photon Propagation and Causality}
\vskip0.5cm
We are concerned with photon propagation in QED in a fixed
background curved spacetime. In this paper,
we consider the properties of photon propagation implied by the
effective action in the form derived by Drummond and Hathrell[1],
$$
\C ~~=~~\int dx \sqrt{-g} \biggl[ -{1\over4}F_{\m\n} F^{\m\n}
+ {1\over m^2}\biggl( a R F_{\m\n} F^{\m\n}
+ b R_{\m\n} F^{\m\l} F^{\n}{}_{\l}
+ c R_{\m\n\l\r} F^{\m\n} F^{\l\r} \biggr) \biggr]
\eqno(2.1)
$$
Here, $a = -{1\over144}{\a\over\pi}$, $b = {13\over360}{\a\over\pi}$
and $c = - {1\over360}{\a\over\pi}$, where $\a$ is the fine structure
constant, and $m$ is the electron mass.
Further contributions to the effective action involving higher
derivatives will be discussed in ref.[10], where we consider the
question of dispersion.
Alternatively, the action (2.1) could be regarded as a starting
point in its own right, with no prior assumption on the magnitude of
the constants $a$, $b$ and $c$ or the scale $m^2$.

The simplest way to determine the characteristics of photon
propagation starting from the effective action is to use geometric
optics. In the leading geometric optics approximation,
the electromagnetic field strength is written as the product of a
slowly varying amplitude and a rapidly varying phase, i.e.
$$
F_{\m\n} = f_{\m\n} e^{i\theta}
\eqno(2.2)
$$
and the wave vector is defined as $k_\m = \partial_\m \theta$. In the
quantum interpretation in terms of photons, $k_\m$ is identified as
the photon momentum. The Bianchi identity,
$$
D_\m F_{\n\l} + D_\n F_{\l\m} + D_\l F_{\m\n} = 0
\eqno(2.3)
$$
becomes
$$
k_\m f_{\n\l} + k_\n f_{\l\m} + k_\l f_{\m\n} = 0
\eqno(2.4)
$$
and constrains $f_{\m\n}$ to be of the form
$$
f_{\m\n} = k_\m a_\n - k_\n a_\m
.\eqno(2.5)
$$
The direction of $a^\m$ specifies the polarisation. Clearly we can
assume $k^\m a_\m = 0$. Of the three remaining possibilities,
only the two orthogonal to the photon momentum are physical
in the quantum theory.

We illustrate this method first for the classical electromagnetic
action. The photon equation of motion is simply
$$
D_\m F^{\m\n} = 0
\eqno(2.6)
$$
i.e.
$$
k_\m f^{\m\n} = 0
\eqno(2.7)
$$
It now follows from eqs.(2.4) and (2.7) that photon trajectories
are null geodesics.
To see this, first multiply (2.7) by $k^\l$ and use the Bianchi
identity. This gives
$$\eqalignno{
0~&=~k_\m k^\l f^{\m\n} \cr
&=~k^2 f^{\l\n} + k^\n k_\m f^{\m\l} \cr
&=~k^2 f^{\l\n}
&(2.8) \cr }
$$
from which we deduce $k^2=0$, i.e. $k_\m$ is a null vector.
Next, it follows from the definition of $k_\m$ as a gradient
that $D_\m k_\n = D_\n k_\m$ and so
$$
k^\m D_\m k^\n ~=~ k^\m D^\n k_\m ~=~ {1\over2} D^\n k^2 ~=~ 0
.\eqno(2.9)
$$
Light rays (photon trajectories) are defined as the integral
curves of the wave vector (photon momentum), i.e. the curves
$x_\m(s)$ for which ${dx_\m \over ds} = k_\m$. Substituting into
eq.(2.9) gives
$$\eqalignno{
0 ~&=~ k^\m D_\m k^\n \cr
&=~ {d^2 x^\n \over ds^2} + \C^\n_{\m\l} {dx^\m \over ds}
{dx^\l \over ds}
.&(2.10) \cr }
$$
This is the geodesic equation.

These familiar properties are no longer true when we consider
the equations of motion derived from the effective action
including the quantum corrections. The Bianchi identity remains
unchanged, but the equation of motion becomes
$$
D_\m F^{\m\n} ~+~ {1\over m^2} \biggl[2b R^\m{}_\l D_\m F^{\l\n}
+ 4c R^{\m\n}{}_{\l\r} D_\m F^{\l\r} \biggr] ~~=~~0
\eqno(2.11)
$$
i.e.
$$
k_\m f^{\m\n} ~+~ {1\over m^2} \biggl[2b R^\m{}_\l k_\m f^{\l\n}
+ 4c R^{\m\n}{}_{\l\r} k_\m f^{\l\r} \biggr] ~~=~~0
\eqno(2.12)
$$
In writing eq.(2.11), we have neglected a number of terms of sub-leading
order:-

\noindent (i)~~Assuming the background gravitational field varies with
the typical curvature scale $L$, terms involving derivatives of the
curvature are suppressed relative to the leading correction by
$O(\l/L)$, where $\l$ is the photon wavelength.

\noindent(ii)~~The Ricci scalar term simply gives a correction to the
$D_\m F^{\m\n}$ coefficient proportional to $a R/m^2$. This is
suppressed by $O(\l_c^2/L^2)$ in the weak field approximation
where we neglect higher powers of the curvature in the
effective action. A term involving the Ricci tensor multiplied
by $D_\m F^{\l\m}$ is neglected for a similar reason.

\noindent (iii)~The standard geometric optics approximation is made
where we neglect derivatives acting on $f_{\m\n}$ relative to those
acting on the phase factor to produce powers of momentum.

Rewriting eq.(2.12) as an equation for the polarisation vector $a^\m$
(which from now on we take to be spacelike normalised,
$a^\m a_\m = -1$), we find
$$
k^2 a^\n ~+~ {2b\over m^2}\Bigl[ R^\m{}_\l \bigl(k_\m k^\l a^\n
- k_\m k^\n a^\l \bigr) \Bigr]
{}~+~ {8c\over m^2} \Bigl[ R^{\m\n}{}_{\l\r} k_\m k^\l a^\r \Bigr]
{}~~=~~0
\eqno(2.13)
$$
In what follows we will be concerned with the modifications to the
light cone condition in a local inertial frame at each point
in spacetime. Introducing local Lorentz components using the
vierbeins $e^a{}_\m$ defined by $g_{\m\n} = \h_{ab} e^a{}_\m e^b{}_\n$,
we have
$$
k^2 a^b ~+~ {2b\over m^2} \Bigl[ R_{ac} \bigl(k^a k^c a^b
- k^a k^b a^c \bigr) \Bigr] ~+~{8c\over m^2} \Bigl[ R_{abcd}
k^a k^c a^d \Bigr] ~~=~~0
\eqno(2.14)
$$
Provided we find a polarisation for which this equation is satisfied,
the corresponding light cone condition is found by contracting
with $a_b$, giving
$$
k^2 ~+~{2b\over m^2} R_{ac} k^a k^c ~-~{8c\over m^2} R_{abcd} k^a k^c
a^b a^d ~~=~~0
\eqno(2.15)
$$

Eqs.(2.14) and (2.15) are manifestly local Lorentz invariant,
provided the curvatures
$R_{abcd}$ and $R_{ab}$ are appropriately transformed. However, the
presence of the (position-dependent) curvature terms means that it does
not reduce to the special relativistic equation at the origin of the LIF
and is different for LIFs at different points in spacetime.
This dynamical equation therefore violates the Principle of
Equivalence, which asserts that all LIFs are equivalent.

\vskip0.3cm
At this point, we should perhaps digress a little on the r\^ole of
the Principle of Equivalence in general relativity. Throughout the
paper, we understand this to refer to the strong Principle of
Equivalence. The so-called weak Principle of Equivalence states that
at each point in spacetime there exists a local Minkowski frame.
This is simply the requirement, fundamental to general relativity, that
spacetime is (pseudo-)Riemannian. The strong Principle of Equivalence,
however, goes on to assert the equivalence of the laws of physics in LIFs
established at different points in spacetime, and furthermore
that these laws take their special relativistic form at the origin
of each LIF. The strong Principle of Equivalence is therefore
an extra arbitrary dynamical condition added to the basic structure
of general relativity and is essentially a specification of minimal
coupling to the gravitational field in the effective action.\footnote
{$^*$}{\eightpoint
\noindent Notice, however, that we do
impose minimal coupling in the classical action in the path integral.
The violation of the Principle of Equivalence occurs at the
quantum level.}
Its r\^ole is merely to exclude
direct curvature couplings such as those appearing in eq.(2.1).

Viewed in this light, there is nothing fundamental about the
Principle of Equivalence. Whether or not it is true is an
experimental question. In the light of the Drummond-Hathrell
result, it seems extremely unlikely that it should remain true,
unless of course some unified theory incorporating QED and the
standard model provided a mechanism for the systematic cancellation
of the vacuum polarisation corrections identified in ref.[1].
In any case, it would be interesting experimentally to look for
evidence of equivalence principle violating interactions such as
those in eq.(2.1) without theoretical prejudice as to the controlling
mass scale (in our case, the electron mass). In principle, such
effects could be observed through, for example, polarisation
dependence in gravitational lensing.

Of course, one apparent motivation for imposing the Principle of
Equivalence is that by excluding curvature couplings {\it a priori},
photons are constrained to follow null geodesics and the question
of possible causality violation does not arise. What Drummond and
Hathrell have shown is that this is not possible in QED.
We shall return to the r\^ole of the Principle of Equivalence
in relation to causality below.

\vskip0.3cm
The photon trajectories corresponding to the new equation of motion
are easily found by a straightforward generalisation of eq.(2.9),
though the resulting equation appears too complicated to be useful
in general. The trajectories satisfy
$$
{d^2 x^\n\over ds^2} ~+~ \C^\n_{\l\r} {dx^\l\over ds} {dx^\r\over ds}
{}~+~ {1\over m^2} D_\n \biggl[ \Bigl(b R_{\l\r} - 4c R_{\l\s\r\t}
a^\s a^\t \Bigr) {dx^\l\over ds} {dx^\r\over ds} \biggr] ~~=~~0
\eqno(2.16)
$$
where $a^\s$ are polarisations satisfying the equation of motion
(2.14).

This brings us to the issue of possible violations of causality due
to the modified trajectories. First, consider the situation in
special relativity. It is certainly true that a spacelike
signal from $A$ to $B$ necessarily corresponds to motion
backwards in time in a class of inertial frames.
However, it should be recognised that this in itself is
not a problem with regard to causality[1]. A causal paradox only arises
if a signal can then be sent back from $B$ to a point $C$ on the
past world line of $A$. The point is that in special relativity,
Poincar\'e invariance (the equivalence of the laws of physics in
all inertial frames) assures us that such a signal is possible,
given the possibility of $A \rightarrow B$. {\it Two} conditions
are necessary to establish a causal paradox -- spacelike motion
{\it and} Poincar\'e invariance.

In the general relativistic context, we of course lose global
Poincar\'e invariance. The nearest analogue to the second condition
above is the Principle of Equivalence, i.e. the equivalence of
the laws of physics in all {\it local} inertial frames. But this is
precisely the condition which we have shown is violated in
establishing the possibility of spacelike photon propagation.
The allowed photon motion depends explicitly on the local curvature.
Provided the Principle of Equivalence is violated, therefore,
it does {\it not} follow that spacelike motion in general
relativity necessarily implies a violation of causality.

Of course, this argument merely removes the most obvious objection
to the possibility of superluminal propagation. It does not prove
that causality violation does not occur -- to prove this would
require showing that trajectories satisfying (2.16) can never
result in signals returning to the past world line of the emitter.
However, since we have established that there is no necessity for
causality violation, there seems no strong reason to doubt that,
despite the modification to the light cone, QED in curved
spacetime remains a causal theory.

\vskip0.3cm
Finally in this section, notice that the equation of motion (2.14)
only admits solutions for certain choices of the polarisation vector
$a^b$. Classically, other polarisations do not propagate -- the
spacetime is opaque to all but the selected polarisations.
However, this clearly has to be reassessed in the quantum theory
where any superposed state is permissible and where we must take
account of the fact that particle trajectories in the above sense
are not well-defined. Suppose, therefore, that we have a polarisation
state which does not correspond to one of the classically allowed
states. To determine how it propagates, we first reexpress it as
a linear superposition of the two permitted polarisation states
transverse to the photon momentum. These states propagate according to
their respective light cone conditions with different $k^2$.
The finally observed state is then in general a different
linear superposition, in the usual way for elementary two-state
quantum mechanical systems.

\vfill\eject

\noindent{\bf 3. The Horizon Theorem and Polarisation Sum Rule}

\vskip0.5cm
Our experience with the explicit examples of photon propagation
in the different types of black hole spacetimes, Schwarzschild[1],
Reissner-Nordstr\"om[2] and Kerr[4], suggests two
general features. First, no matter what happens inside or outside
the horizon, it seems to be a general result that the velocity of
radially directed photons remains equal to $c$ exactly at the
horizon. Second, for the Ricci flat spacetimes, the velocity shifts
for the two transverse polarisations are always equal and opposite.
This is no longer true for non Ricci flat spacetimes such
as FRW [1] and Reissner-Nordstr\"om[2]. In these cases, the
polarisation averaged velocity shift is proportional to the
matter energy-momentum tensor.

These theorems are most easily proved by using the Newman-Penrose
formalism, which characterises spacetimes using a set of complex
scalars found by contracting the Weyl tensor with elements of a
null tetrad.
(For a clear introduction to this formalism, see e.g.~ref.[12].)

For our purposes, we choose the basis vectors of the null tetrad
as follows[13]. Choose $\ell^\m = k^\m$, the photon momentum.
Then, denote the two spacelike, normalised, transverse polarisation
vectors by $a^\m$ and $b^\m$ and construct the null vectors
$m^\m = {1\over\sqrt2}(a^\m + i b^\m)$ and
$\bar m^\m = {1\over\sqrt2}(a^\m - i b^\m)$.
Complete the tetrad with a further null vector $n^\m$ orthogonal
to $m^\m$ and $\bar m^\m$. We therefore have the usual
Newman-Penrose conditions,
$$
\ell.m ~=~ \ell.\bar m ~=~ n.m ~=~ n.\bar m  ~=~ 0
\eqno(3.1)
$$
from orthogonality, and
$$
\ell.\ell ~=~ n.n ~=~ m.m ~=~ \bar m . \bar m ~=~ 0
\eqno(3.2)
$$
since the basis vectors are null. In addition, we impose
$$
\ell.n ~=~1  ~~~~~~~~~~~~~~{\rm and}~~~~~~~~~~~~~~ m.\bar m ~=~ -1
\eqno(3.3)
$$

We denote components with respect to this tetrad frame by the
indices $p,q, \ldots ~=~ 1,2,$ $3,4$ corresponding to
$\ell, n, m, \bar m$ respectively. (We follow the notation
of ref.[12], sect.~1.8.) The metric takes the form
$$
\h^{pq} ~=~ \left(\matrix{0&1&0&0\cr 1&0&0&0\cr
0&0&0&-1\cr 0&0&-1&0\cr}\right)
\eqno(3.4)
$$
In this basis, the components of the Weyl tensor are given in terms
of the Riemann and Ricci tensors by
$$\eqalignno{
C_{pqrs} ~~=~~ R_{pqrs} ~&-~ {1\over2} \bigl(\h_{pr}R_{qs}
- \h_{qr}R_{ps} - \h_{ps}R_{qr} + \h_{qs}R_{pr} \bigr) \cr
&+~  {1\over6} \bigl(\h_{pr}\h_{qs} - \h_{ps}\h_{qr}\bigr) R
&(3.5) \cr }
$$
where $R_{pr} = \h^{qs}R_{pqrs}$ and $R = \h^{pq}R_{pq}$.
The Weyl tensor satisfies the trace-free condition
$$
\h^{ps} C_{pqrs} ~=~ 0
\eqno(3.6)
$$
together with cyclicity
$$
C_{1234} ~+~ C_{1342} ~+~ C_{1423} ~~=~~0
\eqno(3.7)
$$
An important property of the Weyl tensor, which is relevant to our
later discussion, is that it is invariant under conformal (Weyl)
rescalings of the metric.
The ten independent components of the Weyl tensor are denoted by
the five complex Newman-Penrose scalars:
$$\eqalignno{
\Psi_0 ~&=~ - C_{pqrs}\ell^p m^q \ell^r m^s \cr
\Psi_1 ~&=~ - C_{pqrs}\ell^p n^q \ell^r m^s \cr
\Psi_2 ~&=~ - C_{pqrs}\ell^p m^q \bar m^r n^s \cr
\Psi_3 ~&=~ - C_{pqrs}\ell^p n^q \bar m^r n^s \cr
\Psi_4 ~&=~ - C_{pqrs}n^p \bar m^q n^r \bar m^s
&(3.8) \cr }
$$
The components of the Ricci tensor are given a similar
representation[12]. In particular, we will use the notation
$\Phi_{00} = -{1\over2} R_{11} = -{1\over2} R_{pq}\ell^p \ell^q$.

With these preliminaries, we are now ready to state our theorems:-

\vskip0.7cm
\noindent{\bf (a)~~Polarisation Sum Rule}
\vskip0.5cm
\noindent For Ricci flat spacetimes, the sum over the two physical
polarisations of the velocity shift is zero, i.e.
$$
\sum_{\rm pol} \d v ~~=~~ 0
\eqno(3.9)
$$

\noindent For non Ricci flat spacetimes satisfying the Einstein field
equations, we have
$$
\sum_{\rm pol} \d v ~~=~~ -{8\pi\over m^2}(2b+4c)~T_{\m\n} e^\m e^\n
\eqno(3.10)
$$
where $T_{\m\n}$ is the energy-momentum tensor and
$e^\m = k^\m/|\underline k |$ specifies the photon direction.

\vskip0.3cm
If we substitute the vacuum polarisation induced values for
the constants $b$ and $c$ into this equation, we find
$$
\sum_{\rm pol} \d v ~~=~~ - {22\over 45}  {\alpha\over m^2}
{}~T_{\m\n} e^\m e^\n
\eqno(3.11)
$$
Notice the occurrence of the universal coefficient[5] involving the
factor 11. It is straightforward to check that this formula is
consistent with the explicit results for the non Ricci flat FRW [1] and
Reissner-Nordstr\"om[2] spacetimes.

\vskip0.5cm
We present the proof for the Ricci flat case first.
{}From the modified light cone condition (2.15), eq.(3.9) follows
immediately if we can show
$$
\sum_{\rm pol} R_{abcd} k^a k^c a^b a^d ~~=~~0
\eqno(3.12)
$$
i.e.~in the Newman-Penrose tetrad basis,
$$
C_{pqrs} \ell^p \ell^q \bigl(m^q \bar m^s + \bar m^q m^s\bigr)~~=~~0
\eqno(3.13)
$$
Notice that it is legitimate to take $k^a$ to be null in eq.(3.12).
This is consistent with the perturbation expansion in powers of
$O(\l_c^2/L^2)$ used in the starting point (2.15).

Now,
$$
C_{pqrs}\ell^p m^q \ell^r \bar m^s ~=~ C_{1314} ~=~ 0
\eqno(3.14)
$$
from the trace-free condition on the Weyl tensor. To see this,
write out eq.(3.6) explicitly,
$$
C_{1qr2} ~+~ C_{2qr1} ~-~ C_{3qr4} ~-~ C_{4qr3} ~~=~~ 0
,\eqno(3.15)
$$
set $q=r=1$, and use the symmetries of the Weyl tensor.
This establishes (3.13).

\vskip0.3cm
Now consider the non-Ricci flat case. Here,
$$
\sum_{\rm pol} R_{abcd} k^a k^c a^b a^d ~~=~~
C_{pqrs} \ell^p \ell^r \bigl(m^q \bar m^s + \bar m^q m^s\bigr)
{}~~-~~ R_{pr} \ell^p \ell^r
\eqno(3.16)
$$
{}From eq.(2.15) we therefore have
$$
\sum_{\rm pol} k^2 ~~=~~ -{1\over m^2}(4b+8c)~R_{ac} k^a k^c
\eqno(3.17)
$$
For spacetimes satisfying the Einstein field equations we can
simply replace $R_{ac}$ by $8\pi T_{ac}$ (since $k^a$ is null)
and eq.(3.10) follows.

\vskip0.3cm
This brings us to the second theorem, which applies specifically
to black hole spacetimes with an event horizon.

\vskip0.5cm
\noindent{\bf (b)~~Horizon Theorem}
\vskip0.3cm
\noindent At the event horizon, photons with momentum directed
normal to the horizon have velocity equal to $c$, i.e.~the
light cone remains $k^2=0$, independent of their
polarisation.\footnote{$^*$}{\eightpoint
\noindent This theorem has been formulated and proved independently
by G.W. Gibbons and by M.J. Perry (private communication)[16]. }

\vskip0.3cm
We prove this in general for Ricci flat or non Ricci flat spacetimes.
The null tetrad is chosen as above, so that the physical, spacelike,
polarisation vectors $a^\m$ and $b^\m$ lie in the event horizon 2-surface
while $k^\m$ is the null vector normal to them. From eq.(2.15)
we have
$$\eqalignno{
k^2 ~~&=~~ -{2b\over m^2} R_{ac} k^a k^c ~+~ {8c\over m^2}
R_{abcd} k^a k^c a^b a^d \cr
&=~~ -{1\over m^2} (2b+4c) R_{pr} \ell^p \ell^r
{}~+~ {4c\over m^2} C_{pqrs} \ell^p \ell^r (m+\bar m)^q (m+\bar m)^s
&(3.18) \cr }
$$
in the tetrad basis. Using eq.(3.13), and assuming the Ricci tensor
is given in terms of the energy-momentum tensor by the Einstein
field equations, this reduces to
$$
k^2 ~~=~~ -{8\pi\over m^2} (2b + 4c) T_{pr} \ell^p \ell^r
{}~+~ {4c\over m^2} \bigl(C_{pqrs} \ell^p \ell^r m^q m^s ~+~ {\rm c.c.}
\bigr)
\eqno(3.19)
$$

In general, this is non-zero. However, at the event horizon itself,
both the Ricci tensor term
$\Phi_{00} = -4\pi T_{pr} \ell^p \ell^r$ and the Newman-Penrose scalar
$\Psi_0 = C_{pqrs} \ell^p m^q \ell^r m^s$ are zero
for stationary spacetimes.\footnote{$^*$}{\eightpoint
\noindent See ref.[13] for a careful discussion, including the
distinction for non-stationary spacetimes between the event horizon
and the apparent horizon. }

The proof of these assertions is not simple and may be found in
lectures by Hawking[13] (see also ref.[14]).
They follow from consideration of the convergence and shear
of the generators of the event horizon.
The physical interpretation, on the other hand, is clear --
the Ricci term represents the flow of matter across the event horizon
while the Weyl term represents the flow of gravitational radiation
across the event horizon. Both are zero in the classical theory.

This establishes $k^2=0$ at the level of weak field perturbation theory
at which we are working.
It is tempting to speculate that the result is more general and
would hold also for strong gravitational fields, where we do not
have an explicit expression for the modified light cone.
The theorem ensures that the geometric event horizon remains
a true horizon for real photons in QED.

\vskip0.5cm
Before leaving this section, we make a final comment on the
light cone condition for photons propagating in a weak gravitational
field. We have shown that for an individual polarisation state,
$$\eqalignno{
k^2 ~~&=~~ {1\over m^2} (4b+8c) \Phi_{00} ~\pm~
{4c\over m^2} \bigl( \Psi_0 + \Psi_0^* \bigr)  \cr
&=~~ -{8\pi\over m^2}(2b+4c)~T_{\m\n} k^\m k^\n  ~\pm~
{4c\over m^2} \bigl( \Psi_0 + \Psi_0^* \bigr)
&(3.20) \cr }
$$
while the polarisation sum removes the Weyl term, leaving
$$
\sum_{\rm pol} k^2 ~~=~~- {8\pi\over m^2} (4b+8c)~T_{\m\n} k^\m k^\n
\eqno(3.21)
$$
The weak energy condition[14] in gravitation theory implies
$T_{\m\n} k^\m k^\n \ge 0$ for any null vector $k^\m$. Assuming
this to be true, we always have $\sum_{\rm pol} \d v \le 0$.
The relation between the polarisation summed velocity shift and
the matter energy-momentum tensor is consistent with, but
more general than, the observation of Latorre et al.[5]
and should be compared with the corresponding result for photon
propagation in a background electromagnetic field in section 4.
The specifically gravitational birefringent shift in the photon
velocity dependent on the Weyl curvature shows up in the
second contribution to eq.(3.20) proportional to the Newman-Penrose
scalar $\Psi_0$.

\vfill\eject

\noindent{\bf 4. Electromagnetic Birefringence}
\vskip0.3cm
In an earlier paper[2], we calculated the modification to the velocity
of light in an arbitrary anisotropic (but homogeneous) electromagnetic
field in flat spacetime. In this section, we discuss these results a
little further, first to make contact with the formula of
Latorre et al.[5] and the gravitational formulae in section 3,
and second to discuss the relation with the conformal anomaly.

The starting point is the Euler-Heisenberg effective action,
$$
\C ~~=~~ \int dx \biggl[ -{1\over 4} F_{ab} F^{ab}   ~+~
{1\over m^4}\biggl(z \Bigl(F_{ab} F^{ab}\Bigr)^2
+ y F_{ab} F_{cd} F^{ac} F^{bd} \biggr)~\biggr]
\eqno(4.1)
$$
where $z = -{1\over 36}\a^2$ and $y ={7\over 90} \a^2$,
from which we derive the equation of motion
$$
D_a F^{ab} ~-~ {16z\over m^4}F^{ab} F_{cd} D_a F^{cd}
{}~-~ {8y\over m^4} \Bigl( F^{ac} F_{cd} D_a F^{bd}
+ F^{ac} F^{bd} D_a F_{cd} \Bigr)  ~~=~~ 0
\eqno(4.2)
$$
These expressions are the analogues of eqs.(2.1) and (2.11) in the
gravitational case and are derived using similar approximations.
In the geometric optics approximation, the equation of motion
becomes[2]
$$
k_a f^{ab} ~-~ {16z\over m^4} F^{ab}F_{cd} k_a f^{cd}
{}~-~ {8y\over m^4} \Bigl( F^{ac}F_{cd} k_a f^{bd}
+ F^{ac}F^{bd} k_a f_{cd} \Bigr) ~~=~~ 0
\eqno(4.3)
$$
where $F_{ab}$ is the background electromagnetic field strength.
Using the Bianchi identity to set $f_{ab} = k_a a_b - k_b a_a$,
and rewriting in terms of the polarisation vector $a^a$, we find
$$
k^2 a^b ~-~ {8\over m^4}(4z+y) F_a{}^b F_{cd} k^a k^c a^d
{}~-~ {8y\over m^4}   F_a{}^c F_{cd} \bigl(k^ak^ba^d - k^ak^da^b\bigr)
{}~~=~~ 0
\eqno(4.4)
$$
Provided we have a polarisation satisfying this equation, the
corresponding modified light cone condition is
$$
k^2 ~-~ {8\over m^4}(4z+y) F_{ab}F_{cd} k^ak^ca^ba^d
{}~+~ {8y\over m^4}   F_a{}^d F_{cd} k^ak^c ~~=~~ 0
\eqno(4.5)
$$

It is easy to see that a polarisation satisfying $F_{ab}a^b = 0$ solves
eq.(4.4).\footnote{$^*$}{\eightpoint
\noindent For the case of a pure magnetic field[6], the first and
second polarisation states considered here correspond to
polarisations respectively orthogonal to and coplanar with the
plane spanned by the photon momentum and magnetic field directions. }
The light cone condition is then
$$
k^2 ~~=~~ {-8y\over m^4}    F_a{}^d F_{cd} k^a k^c
\eqno(4.6)
$$
Recalling the form of the classical electromagnetic energy-momentum
tensor,
$$
T_{ac} ~=~ \Bigl(F_a{}^d F_{cd} - {1\over 4}\h_{ac} F_{bd}F^{bd}\Bigr)
\eqno(4.7)
$$
and remembering that consistent with the weak field perturbative
expansion we can take $k^a$ to be null on the r.h.s. of eq.(4.6),
we have
$$
k^2 ~~=~~ -{8y\over m^4}  ~ T_{ac} k^a k^c
\eqno(4.8)
$$
corresponding to a velocity shift for this polarisation state of
$$
\d v ~~=~~ -{4y\over m^4}  ~ T_{ac} e^a e^c
\eqno(4.9)
$$
where as before $e^a = k^a/|\underline k|$.

To deduce the light cone condition for the second polarisation
state, we use the identity
$$
\sum_{\rm pol} a^b a^d ~=~ -\bigl( \h^{bd} - e^b \bar e^d\bigr)
\eqno(4.10)
$$
(where $\bar e^a$ is simply $e^a$ with the sign of the spacelike
components reversed) to show that the two terms in eq.(4.5)
are equal. So, for the second polarisation, we immediately find
$$
k^2 ~~=~~ -{8\over m^4} (4z+2y)~T_{ac} k^a k^c
\eqno(4.11)
$$
that is,
$$
\d v ~~=~~ -{4\over m^4} (4z+2y)~ T_{ac} e^a e^c
\eqno(4.12)
$$

The dependence of the velocity shifts on the energy-momentum tensor
is therefore exactly the same as in the gravitational case, except
that here the polarisation dependence enters already in determining
the coefficient of this term. Unlike the gravitational case, therefore,
birefringence occurs here already at the level of the
energy-momentum tensor dependence.
If we now take the polarisation sum, and substitute the explicit values
for the coefficients $y$ and $z$, we find
$$
\sum_{\rm pol} \d v ~~=~~ -{4\over m^4}(4z+3y)~T_{ac}e^a e^c
{}~~=~~ -{22\over 45} {\a^2\over m^4}~ T_{ac} e^a e^c
\eqno(4.13)
$$
Again notice the appearance of the universal coefficient involving
the number 11, just as in the gravitational case.

Writing out the energy-momentum tensor in terms of $\underline E$
and $\underline B$ fields, we find
$$
T_{ac} e^a e^c ~~=~~ \underline E^2 + \underline B^2 -
(\underline E . \underline n)^2 - (\underline B . \underline n)^2
- 2 (\underline E \times \underline B) . \underline n
\eqno(4.14)
$$
where $\underline n$ is the direction of the photon momentum.
In this form, we recover the results derived by a less direct method
in ref.[2] (in the notation used there, $X = T_{ac} k^a k^c$).
It was pointed out there that since the coefficients in both
eqs.(4.9) and (4.12) are negative and $T_{ac} e^a e^c \ge 0$
(the r.h.s. of eq.(4.14) is a positive definite quantity), both
polarisations have velocities less than $c$, as required by
causality in this special relativistic context.

As a final embellishment, we can follow ref.[5] and consider the
direction averaged velocity shift.
It is clear from the expression (4.14) that if we average over
the photon direction, the $\underline E . \underline n$,
$\underline B . \underline n$ and Poynting vector
$(\underline E \times \underline B) . \underline n$ terms disappear.
The remainder is just the energy density of the electromagnetic field.
So we have
$$
\sum _{\rm pol} \langle \d v\rangle ~~=~~ -{4\over m^2} (4z+3y)
\bigl(\underline E^2 + \underline B^2\bigr)
\eqno(4.15)
$$
where $\langle{}\rangle$ denotes direction averaging.
This is in agreement with the observation of ref.[5].

\vskip0.5cm
We now present an intriguing connection between these results and the
conformal anomaly. The Euler-Heisenberg action can be rewritten as
$$
\C ~~=~~ \int dx \biggl[ - {\cal F} ~+~ {1\over m^4}
\Bigl( 8 (2z+y) {\cal F}^2 + 4y {\cal G}^2 \Bigr) ~\biggr]
\eqno(4.16)
$$
in terms of the Lorentz invariants ${\cal F} = {1\over 4} {\rm tr}F^2
= {1\over 2}(\underline E^2 - \underline B^2)$ and
${\cal G} = {1\over 4} {\rm tr} F F^* = - \underline E . \underline B$.

The vacuum expectation value of the energy-momentum tensor for QED
in a background electromagnetic field is given to one loop in terms
of this effective Lagrangian ${\cal L}$ by[15]
$$
\langle T_{ab} \rangle
{}~~=~~ -\Bigl( F_{ac} F_b{}^c - {1\over4} \h_{ab}F^2
\Bigr){\partial {\cal L}\over \partial {\cal F}} ~+~
\h_{ab} \biggl({\cal L} - {\cal F}{\partial {\cal L}\over
\partial {\cal F}} - {\cal G}{\partial {\cal L}\over \partial{\cal G}}
\biggr)
\eqno(4.17)
$$
and so the conformal anomaly is
$$\eqalignno{
\langle T^a{}_a \rangle ~~&=~~ 4 \biggl({\cal L} - {\cal F}
{\partial {\cal L}\over \partial {\cal F}} - {\cal G}
{\partial {\cal L}\over \partial {\cal G}} \biggr) \cr
&=~~ -{16\over m^4} \Bigl(~(4z+2y) {\cal F}^2 ~+~ y {\cal G}^2~\Bigr)
&(4.18) \cr }
$$
Now compare with the velocity shift formulae (4.9) and (4.12).
Notice that the coefficients of the ${\cal F}^2$ and ${\cal G}^2$
terms in the conformal anomaly are precisely those appearing in the
velocity shifts for the two polarisations.

The requirement that both velocity shifts are negative is therefore
equivalent to the requirement that the VEV of the trace of the
energy-momentum tensor in the background electromagnetic field
is negative. The sign of the conformal anomaly is therefore
linked to causality in photon propagation. This is certainly a
curious result, but beyond noting that it is one more hint of a
deeper connection between photon propagation with a modified
light cone and the conformal anomaly, we have no real physical
understanding of why it should be true.

\vskip1.5cm

\noindent{\bf Acknowledgements}
\vskip0.5cm
I would particularly like to thank Ian Drummond and Gary Gibbons
for discussions, correspondence and for making a number of
suggestions which have been pursued in this paper.
Useful discussions with R. Daniels, S. Deser, H. Osborn,
K. Scharnhorst, R. Tarrach and G. Veneziano are also gratefully
acknowledged.

\vfill\eject

\noindent{\bf References}
\vskip0.5cm

\settabs\+\ [&1] &G.M.Shore  \cr

\+\ [&1] &I.T. Drummond and S.J. Hathrell, Phys. Rev. D22 (1980)
343 \cr
\+\ [&2] &R.D. Daniels and G.M. Shore, Nucl. Phys. B425 (1994) 634 \cr
\+\ [&3] &Y. Ohkuwa, Prog. Theor. Phys. 65 (1981) 1058 \cr
\+\ [&4] &R.D. Daniels and G.M. Shore, `Faster than light photons
and rotating black holes', \cr
\+\ &{}  &\hskip1cm Swansea preprint SWAT 95/71, in preparation \cr
\+\ [&5] &J.L. Latorre, P. Pascual and R. Tarrach,
Nucl. Phys. B437 (1995) 60 \cr
\+\ [&6] &S.L. Adler, Ann. Phys. (N.Y.) 67 (1971) 599 \cr
\+\ [&7] &K. Scharnhorst, Phys. Lett. B236 (1990) 354 \cr
\+\ [&8] &G. Barton, Phys. Lett. B237 (1990) 559 \cr
\+\ [&9] &G. Barton and K. Scharnhorst, Univ.~of
Sussex preprint 9242 (1992)\cr
\+ [1&0] &R.D. Daniels and G.M. Shore, `Faster than light photons
in gravitational fields -- \cr
\+\ &{}  &\hskip1cm dispersion and the effective action',
Swansea preprint SWAT 95/55, in prep. \cr
\+ [1&1] &S.W. Hawking,  Comm. Math. Phys. 43 (1975) 199 \cr
\+ [1&2] &S. Chandrasekhar, `The mathematical theory of black holes',\cr
\+\ &{}  &\hskip1cm Oxford University Press, 1983 \cr
\+ [1&3] &S.W. Hawking, `The event horizon', 1972 Les Houches
lectures, ed. B. de Witt, \cr
\+\ &{}  &\hskip1cm Gordon and Breach, 1972 \cr
\+ [1&4] &S.W. Hawking and G.F.R. Ellis, `The large scale structure
of space-time', \cr
\+\ &{}  &\hskip1cm Cambridge University Press, 1973 \cr
\+ [1&5] &J. Schwinger, Phys. Rev. 82 (1951) 664 \cr
\+ [1&6] &G.W. Gibbons and M.J. Perry, to be published \cr

\bye